# A new strategy to conceal an object from electromagnetic wave

Yu Luo, Jingjing Zhang, Hongsheng Chen, Bae-Ian Wu, and Jin Au Kong

*The Electromagnetics Academy at Zhejiang University, Zhejiang University, Hangzhou 310058, P. R. China and Research Laboratory of Electronics, Massachusetts Institute of Technology, Cambridge, Massachusetts 02139, USA*

A new recipe for concealing objects from detection is suggested. Different with traditional cloak which deflects light around the core of the cloak to make the object inside invisible, our cloak guides the light to *penetrate* the core of the cloak but without striking some region of the cloak shell - the so called 'folded region'. Full wave analytical calculation shows that this cloak will lead to a scattering enhancement instead of scattering reduction in contrast to the traditional cloak; the scattered field distribution can also be changed as if the scatterer is moved from its original position. Such interesting phenomenon indicates the proposed cloak can be used to disguise the true information of the object, e.g. the position, the size, etc, and further mislead the observer and avoid being detected.

How to conceal an obstacle from electromagnetic wave detection has been a hot research topic. When talking about "invisible" of an object, we originally mean we cannot detect the object by receiving the reflected light with our eyes. Therefore, most of the cloak designs proposed so far are trying to make the Radar Cross Section (RCS) of the objects as small as possible. For example, Alù and Engheta presented the method of using plasmonic and metamaterial covers to reduce the total scattering of a single or a collection of spheres, which relys on an inherently nonresonant scattering cancellation phenomenon [1, 2]; Cloaking device based on anomalous localized resonance, the cylindrical superlens, was proposed by Milton *et. al.* [3, 4]; The transformation approach [5-20] to design zero RCS cloak has recently won even more concerns since the yielded cloaking device is insensitive to the object being masked.

In this paper, a new recipe for concealing objects from detection is suggested. The purpose of the proposed cloak is to disguise the true information of the object, e.g. the position, the size, etc, and further mislead the observer and avoid being detected. Different from the conventional cloak which excludes the light from the central region, our cloak squeezes the bulk of incident light into a narrow beam that penetrates the core but without touching some regions of the outer layer. Full wave analytical calculation shows that the cloak can enhance the scattering from the obstacle so that it seems to be produced by a much larger scatterer, or 'move' the scattered field as if the object is moved from its original position. Such kind of cloak can be implemented by isotropic negative index materials with the magnitude of the spatially varying refractive index larger than one everywhere. Under this condition, concealment can be achieved with artificially structured negative-index media (NIM) like phonic crystal [21-23] or subwavelength dielectric resonator [24,25].

As prescribed in Ref [5], suppose the spherical cloak is an annulus with inner radius $R_1$ and outer radius $R_2$, under a radial mapping $r' = f(r)$, the parameters of the three-dimensional transformation media will take the form [18]

$$\varepsilon_r = \mu_r = \frac{f^2(r)}{r^2 f'(r)}, \quad \varepsilon_\theta = \mu_\theta = \varepsilon_\varphi = \mu_\varphi = f'(r) \tag{1}$$

The invisibility property requires $f(R_1) = 0$. Under this condition, it can be easily concluded from Equation (1) that the parameters are strongly anisotropic and the radial components are even zero at the inner surface. In order to reduce the anisotropy of the cloak, we choose $f(r)$ to be $f_1(r) = \frac{R_2^2}{r}$, which is monotone decreasing as depicted in Fig. 1(a) ($f(r) = r$ represents the free space in the outer region). The corresponding permittivity and permeability become

$$\varepsilon_1 = \mu_1 = -\frac{R_2^2}{r^2} \tag{2}$$

which means the coating is constructed by an isotropic left-handed metamaterial (LHM) and the magnitude of the parameters are all larger than 1 (since $R_1 < r < R_2$).

To begin with, we first study the scattering property when a conductive sphere with radius $R_1$ is coated by this metamaterial shell. Suppose an $E_x$ polarized plane wave with a unit amplitude $E_i = \hat{x} e^{i k_0 z}$ is incident upon the coating along the $z$ direction. Following the same steps of Ref [13,18], the scattered coefficients $b_n^{TM}$ and $b_n^{TE}$ for the scattered wave can be obtained by using Mie scattering theory

$$b_n^{TM} = -a_n \frac{\psi_n'\left(k_0 R_2^2 / R_1\right)}{\zeta_n'\left(k_0 R_2^2 / R_1\right)}, \quad b_n^{TE} = -a_n \frac{\psi_n\left(k_0 R_2^2 / R_1\right)}{\zeta_n\left(k_0 R_2^2 / R_1\right)} \tag{3}$$

where $a_n = \frac{(-i)^{-n}(2n+1)}{n(n+1)}$, $n = 1, 2, 3, \ldots$; $\psi_n(\xi)$, $\zeta_n(\xi)$ represent the Riccati-

Bessel function of the first and the third kind, respectively [27]. Equations (3) directly show the scattered coefficients are the same as that of the case when a conductive sphere with radius $R_2^2/R_1$ is placed in free space [28], indicating that an obstacle with smaller radius $R_1$ will cause larger scattering ($R_2^2/R_1$ is larger). Fig. 1(b) shows the calculated electric fields in this case. The inner and the outer radius of the coating are $R_1 = 5\,cm$ and $R_2 = 10\,cm$, respectively. The frequency of the plane wave is set to $2GHz$, corresponding to a wavelength of $15cm$. The total field distribution of the same plane wave incident upon a conductive sphere with radius $20\,cm$ placed in free space is depicted in Fig. 1(c). It is seen from Fig. 1(b) and Fig. 1(c), the total field distributions are exactly the same in the region $r > R_2^2/R_1$. In other words, the observer will think the scattereing is caused by the larger obstacle. A more interesting phenomenon which can be seen from Fig. 1(b) is that the total field is very large in the region $R_1 < r < R$. This large response represents accumulation of energy over many cycles, such that a considerable amount of energy is stored in region relative to the driving field. Further physical insight will be related in the later part of this paper.

In order to hide the position information of an obstacle, we fill the inner region of this LHM coating with another right-handed material (RHM) with the parameters

$$\varepsilon_2 = \mu_2 = \left(\frac{R_2}{R_1}\right)^2. \qquad (4)$$

In fact, we can take the LHM coating along with the RHM core as a whole concealing device, which is described a non-monotonic function shown in Fig. 2 (a). Note that in the folded region $R_0 < r < R_2$, a certain value of $f(r)$ corresponds to two different radii. If we place an object at point **A** in this region of the concealing device, the scattering caused by the object will be identical to that from a larger object (with a size magnification of $\left(\dfrac{R_2}{R_1}\right)^2$) located at point **B** in free space ($r > R_2$), which can be validated by the aforementioned scattering theory. Numerical simulations based on finite element method also verify this idea and the field distributions with a plane wave incidence are depicted in Fig. 2. The frequency of the plane wave is 2 *GHz*. $R_1$ and $R_2$ are 5 *cm* and 10 *cm* respectively. The total field distribution when a conductive sphere with radius 1.5 *cm* is located at point **A** in the folded region is shown in Fig. 2(b), while Fig .2(c) displays the case when a conductive sphere with radius 6 *cm* is placed at point **B** in free space. By comparison, we find that the field distributions in the two cases are identical, indicating that an observers outside will "see" a bigger scatterer (with a size magnification of 4 in this case) which has been shifted from its original position and could not find out its real position, therefore, the volume and position information of the object will be disguised to mislead the observer. In order to understand this phenomenon from physical perspective we consider the case when a plane wave is incident upon the spherical concealing device in the free space, as shown in Fig.2 (d). Similar to the case in Ref. [18], the wave has penetrated into the core, making some energy circulating in the folded region without introducing any scattering, which is because that the whole system will store energy from the incident wave before the steady state is reached Therefore, the scattering shift and magnification phenomena here are also formed in time harmonic state, which is similar to the imaging of perfect lens [26].

    We now use ray tracing exercise to show some interesting phenomena of this type of cloak. When the size of the object to be hidden is much larger than the wave length, geometrical optics approximation can be applied in the calculation. By taking

the geometry limit of Maxwell's equations, the rays of light can be obtained with numerical integration of a set of Hamiltion's equations [5]. The case in the absence of the concealing device is depicted in Fig .3(a). The paths of the rays are parallel to the *z* direction. As shown in this figure, the obstacle will be detected by the wave whose path has been marked in blue dashed line. Fig .3(b) shows the case when the same obstacle is placed inside a traditional cloak proposed in [5]. We see that the light paths have been bent around the center of shell, making the obstacle inside the core invisible. While in Fig .3(c), our isotropic concealing device is utilized to hide the object. Different with perfect cloak in Fig .3(b) which can guide the rays around the central volume, here the lights have penetrated into the core, but are all bent around the folded region. Thereby, the object still cannot be detected as long as it is located in the folded region inside the LHM concealing coating.

It should be pointed out that this phenomenon does not contradict with the full wave results we show in the former part of the letter. In fact, they have the common ground that all the waves incident upon the outer surface of the metamaterial coating have been concentrated in a small region ($r < R_0$, $R_0$ is depicted in Fig .2(a)) as they penetrate into the core (see Fig .3(c) and Fig .2(b)). If we considered the incident beam as a wave, the power will diffract into the folded region due to the high spatial dispersion. Therefore, the energy will be stored there and circulate between the LH coating layer and the RH core. It is also the reason why the power flows inside the core is larger than the power flows through the whole system [18]. However, if spatial dispersion of the incident beam is very small and can be neglected, or if the incident beam can be considered as a ray, the beam will transmit through the concealing device following the paths depicted in Fig .3(c) before they diffract into the folded region.

Another point which should be noted is that the whole device is constructed by isotropic media, thus in geometric optics limit, the effective refraction index is mostly important to the light paths. It is required that the refraction index of the LH coating layer and the RH core are $n_1 = -\frac{R_2^2}{r^2}$ and $n_2 = \left(\frac{R_2}{R_1}\right)^2$ respectively (these parameters are also applicable to 2 dimensional cylindrical case), which is physically realizable with artificially structured metamaterials, such as phonic crystal [21-23] and high dielectric resonators [24,25]. The reflection at the interfaces can be suppressed by controlling the thickness and surface termination of the LH coating layer.

In conclusion, analytical full wave theory and ray tracing method have been introduced to examine the behavior of a concealing device consisting of a spherical LH coating layer and RH core. It is demonstrated that this concealing device will lead to a scattering enhancement instead of scattering reduction in contrast to the traditional cloak; the scattered field distribution can also be changed as if the scatterer is moved from its original position. Our proposed concealing device therefore can be used to disguise the true information of the object, e.g. the position, the size, etc, and further mislead the observer and avoid being detected. Our approach provides another way to achieve concealment, and complement the cloaking devices published.

This work is sponsored by the Chinese National Science Foundation under Grant Nos. 60531020, 60671003 and 60701007, the NCET-07-0750, the ZJNSF (R105253), the Ph.D. Programs Foundation of MEC (No. 20070335120), the ONR under Contract No. N00014-01-1-0713, and the Department of the Air Force under Air Force Contract No.F19628-00-C-0002. We are grateful for the helpful discussions with Prof. Lixin Ran.

**Figure Captions**

FIG 1: (color online) (a) Schematic figure of the transformation functions $f(r)$ of the coated layer. (b) Field distribution due to a $E_x$ polarized plane wave incident upon a conducting sphere with radius $R_1 = 0.05$ $m$ coated metamaterial with $R_2 = 0.1$ $m$. (c) Field distribution due to an $E_x$ polarized plane wave incident upon a conducting sphere with radius $R = 0.2$ $m$ in free space.

FIG 2: (color online) (a) Schematic figure of the functions $f(r)$ which is non-monotonic. The inner media ($0 < r < R_1$) is RHM, while the media in the coating ($R_1 < r < R_2$) is LHM. The fold ($R_0 < r < R_2$) indicates the hidden region. (b) Field distribution of the same plane wave incident upon the coating when a conductive sphere with radius 1.5 *cm* is placed at the position **A** in the hidden region of coating. (c) Field distribution of this plane wave incident upon a conductive sphere with radius 6 *cm* placed at the position **B** in free space. (d) Field distribution of an $E_x$ polarized plane wave incident upon the coating when there are no scatterers inside the coating.

FIG 3: (color online) (a) Ray trajectories in the absence of the concealing device. The blue dashed line indicates that the scatterer can be detected by the incident ray. (b) The path of the rays when the scatterer is placed in an ideal transformation cloak proposed by J. Pendry. No lights can enter into the center of the shell. (c) The path of the rays when the scatterer is located at the hidden region of the coating. The lights have been smoothly bent around this region.

FIG. 1

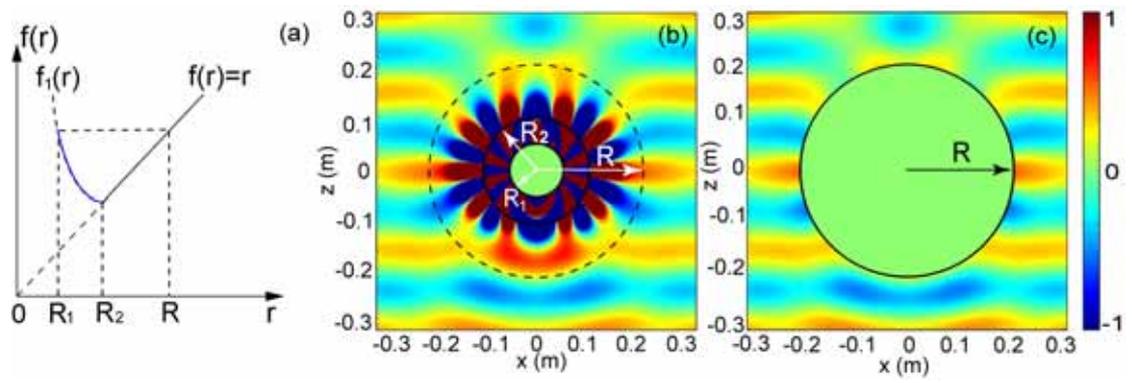

FIG. 2

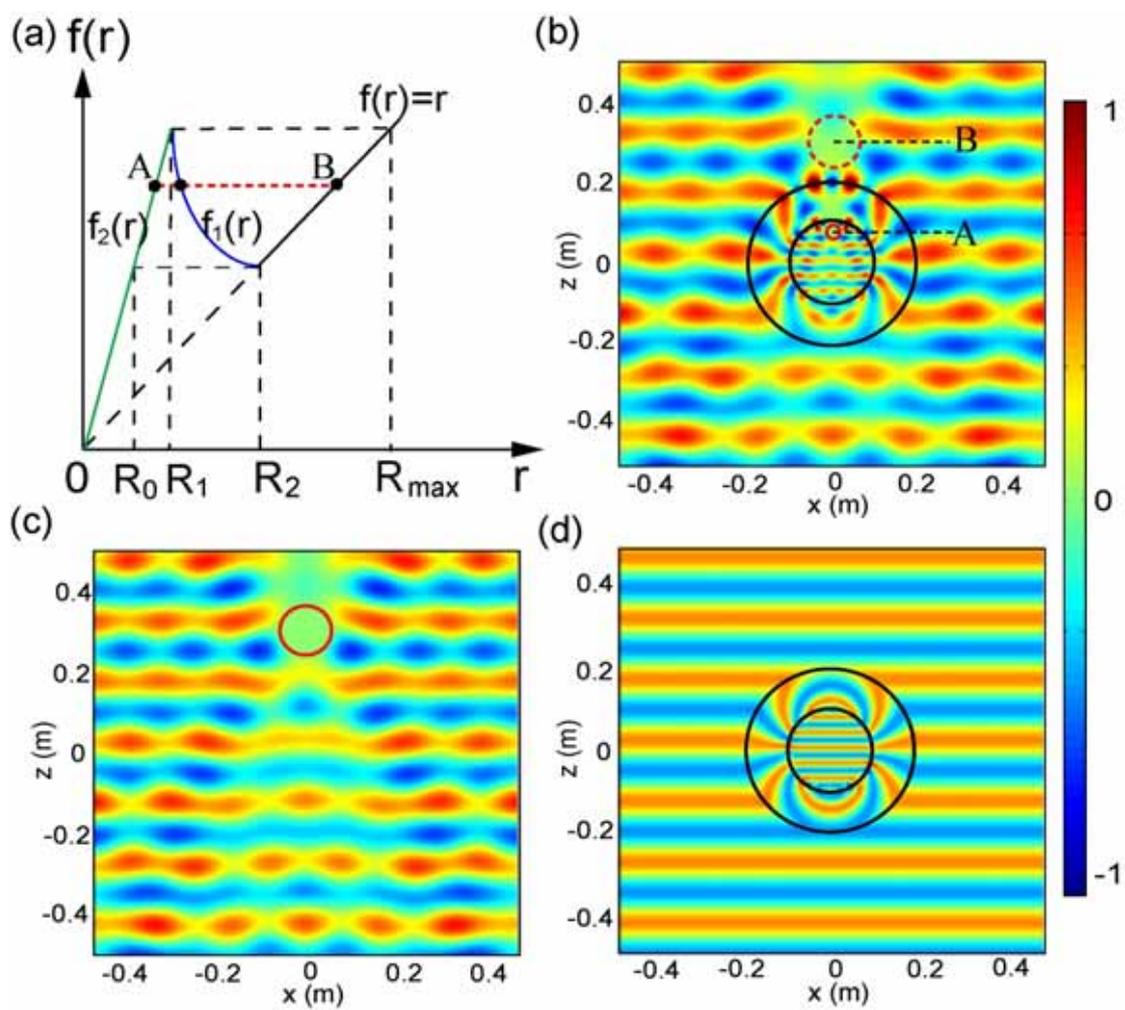

FIG. 3

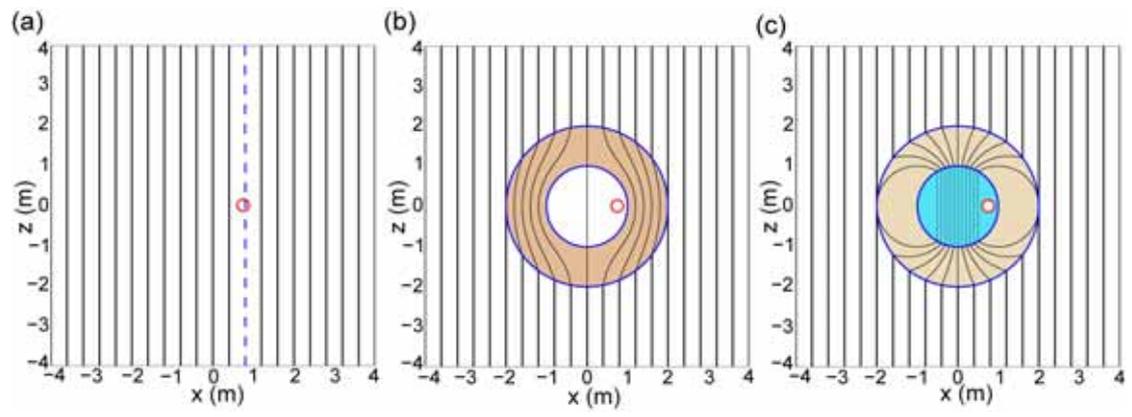